\documentclass[a4paper,fleqn,usenatbib]{mnras}

\usepackage[T1]{fontenc}
\usepackage{ae,aecompl}

\usepackage{graphicx}	% Including figure files
\usepackage{amsmath}	% Advanced maths commands
\usepackage{amssymb}	% Extra maths symbols

\newcommand{\mvir}{M_\mathrm{vir}}
\newcommand{\rvir}{R_\mathrm{vir}}
\newcommand{\rmax}{R_\mathrm{max}}
\newcommand{\vvir}{V_\mathrm{vir}}
\newcommand{\vmax}{V_\mathrm{max}}
\newcommand{\vpeak}{V_\mathrm{peak}}

\newcommand{\lcdm}{$\Lambda$CDM}

\newcommand{\vcirc}{V_{\rm circ}}

\title[Properties of resonant sterile neutrino dark matter subhalos]{Properties of Resonantly Produced Sterile Neutrino Dark Matter Subhalos}

\author[S. Horiuchi et al.]{Shunsaku Horiuchi,$^{1}$\thanks{E-mail: horiuchi@vt.edu}
Brandon Bozek,$^{2}$
Kevork N.~Abazajian$^{3}$
\newauthor
Michael Boylan-Kolchin$^{2}$
James S.~Bullock,$^{3}$
Shea Garrison-Kimmel,$^{4}$
\newauthor
and Jose Onorbe$^{5}$
\\
$^{1}$Center for Neutrino Physics, Department of Physics, Virginia Tech, 850 W Campus Drive, Blacksburg, VA 24061, USA\\ 
$^{2}$Department of Astronomy, The University of Texas at Austin, 2515 Speedway, Stop C1400, Austin, TX 78712, USA \\
$^{3}$Center for Cosmology, Department of Physics and Astronomy, University of California, Irvine, CA 92697, USA \\
$^{4}$TAPIR, California Institute of Technology, Pasadena, CA 91125, USA \\
$^{5}$Max-Planck-Institut fuer Astronomie, Koenigstuhl 17, 69117 Heidelberg, Germany 
}

\date{Accepted XXX. Received YYY; in original form ZZZ}

\pubyear{2015}

\begin{document}
\label{firstpage}
\pagerange{\pageref{firstpage}--\pageref{lastpage}}
\maketitle

% Abstract of the paper
\begin{abstract}
The anomalous 3.55 keV X-ray line recently detected towards a number of massive dark matter objects may be interpreted as the radiative decays of 7.1 keV mass sterile neutrino dark matter. Depending on its parameters, the sterile neutrino can range from cold to warm dark matter with small-scale suppression that differs in form from commonly-adopted thermal warm dark matter. Here, we numerically investigate the subhalo properties for 7.1 keV sterile neutrino dark matter produced via the resonant Shi-Fuller mechanism. Using accurate matter power spectra, we run cosmological zoom-in simulations of a Milky Way-sized halo and explore the abundance of massive subhalos, their radial distributions, and their internal structure. We also simulate the halo with thermal 2.0 keV warm dark matter for comparison and discuss quantitative differences. We find that the resonantly produced sterile neutrino model for the 3.55 keV line provides a good description of structures in the Local Group, including the number of satellite dwarf galaxies and their radial distribution, and largely mitigates the too-big-to-fail problem. Future searches for satellite galaxies by deep surveys, such as the Dark Energy Survey, Large Synoptic Survey Telescope, and Wide Field Infrared Survey Telescope, will be a strong direct test of warm dark matter scenarios.  
\end{abstract}

% Select between one and six entries from the list of approved keywords.
% Don't make up new ones.

%%%%%%%%%%%%%%%%%%%%%%%%%%%%%%%%%%%%%%%%%%%%%%%%%%

%%%%%%%%%%%%%%%%% BODY OF PAPER %%%%%%%%%%%%%%%%%%

\section{Introduction}

The Lambda cold dark matter ($\Lambda$CDM) paradigm has been extremely successful in explaining a variety of observations on cosmological scales \citep[e.g.,][]{2012ApJ...761...14H,2013ApJS..208...19H,2015A&A...580A..22P}. Tests on galactic and sub-galactic scales, while often complicated by the physics of galaxy formation, provide crucial verification of the $\Lambda$CDM model. Several issues with $\Lambda$CDM on small-scales have been highlighted in the past decade, prompting investigations into galaxy formation within the $\Lambda$CDM model, as well as alternative models of DM. 

Among the earliest and most prominent small-scale issue is the ``missing satellites'' problem: the large excess in the number of expected satellite galaxies of the Milky Way (MW), which outnumber those observed by a factor of 10 or more \citep{1999ApJ...522...82K,1999ApJ...524L..19M}. Another issue is the core/cusp problem, which is seen in, e.g., the flatter central dark matter density profiles observed in many low-surface brightness (LSB) galaxies \citep[e.g.,][]{2005ApJ...621..757S,2009MNRAS.397.1169D,2011AJ....141..193O} when compared with the cuspy profiles predicted by pure $\Lambda$CDM simulations \citep{1994ApJ...427L...1F,1997ApJ...490..493N}. More recently, the most massive subhalos of MW-sized hosts have been shown to contain too much dark matter to accommodate the stellar kinematic data of the observed MW dwarf spheroidal (dSph) satellite galaxies \citep{2011MNRAS.415L..40B,2012MNRAS.422.1203B}. The number of massive subhalos failing in this way (``massive failures'') is typically $\sim 20$, depending on the halo mass, cosmology, and other variables \citep[e.g.,][]{2014MNRAS.444..222G,2015arXiv150901255G}, and constitute the ``Too big to Fail'' (TBTF) problem.

Whether these small-scale issues motivate a change to the $\Lambda$CDM paradigm depends quantitatively on the degree to which feedback and other baryonic processes can remedy the discrepancies. For example, studies have postulated how galaxy formation does not follow the subhalo mass at $z=0$, but rather, the maximum mass at some earlier epoch, often at reionization. This reduces the number of expected satellites and helps to remedy the missing satellites problem \citep{2000ApJ...539..517B,2004ApJ...609..482K,2005ApJ...629..259R,2009ApJ...696.2179K,2009MNRAS.399L.174O,2010ApJ...710..408B}. Simultaneously, searches with the Sloan Digital Sky Survey (SDSS) has revealed some dozen more dSph galaxies \citep{2007ApJ...654..897B}, and searches in the southern hemisphere are starting to produce results in the same direction \citep{2015ApJ...807...50B,2015ApJ...805..130K,2015arXiv150803622T}. Studies have also revealed the gross incompleteness of current satellite samples due to a combination of incomplete sky coverage, luminosity bias, and surface brightness limits \citep{2008ApJ...688..277T,2009AJ....137..450W,2014ApJ...795L..13H}. 

Concerning the inner dark matter densities, baryonic processes can effectively reduce the central dark matter densities of LSB host halos, addressing the core/cusp problem \citep[e.g.,][]{2012MNRAS.421.3464P,2012MNRAS.422.1231G,2014MNRAS.441.2986D,2015arXiv150202036O,2015ApJ...809...69S,2015arXiv150201947D}. However, \cite{2012ApJ...759L..42P} and \cite{2013MNRAS.433.3539G} have argued that these effects are insufficient to affect dimmer satellites relevant for TBTF \citep[but, see][]{2013ApJ...765...38G,2014ApJ...782L..39A}. On the other hand, environmental effects, coupled with baryonic interactions, can be conducive to more efficient inner density reduction even on these scales \citep{2012ApJ...761...71Z,2014MNRAS.438.1466A,2014ApJ...786...87B,2014JCAP...04..021D,2015MNRAS.450.3920B,2015MNRAS.454.2092O}. However, the TBTF issue continues to be observed out in field galaxies of the Local Group, where environmental effects are greatly reduced or non-existent \citep{2012MNRAS.425.2817F,2014MNRAS.444..222G,2015A&A...574A.113P}. Therefore, although many plausible mechanisms within the context of $\Lambda$CDM exist to explain small-scale issues, it is not clear if they are where the solution lies. 

Various alternatives to $\Lambda$CDM have been investigated in order to resolve small-scales issues, including non-standard primordial power spectra \citep{2000PhRvL..84.4525K,2002PhRvD..66d3003Z,2014MNRAS.437.2922P,2014MNRAS.444..961G} and modifications to CDM, e.g., self-interacting dark matter \citep[][]{2012MNRAS.423.3740V,2014MNRAS.444.3684V,2013MNRAS.430...81R,2015MNRAS.453...29E}, warm dark matter \citep[][]{2012JCAP...10..047A,2013JCAP...03..014A,2012MNRAS.420.2318L,2014MNRAS.439..300L,2013MNRAS.433.1573S,2014MNRAS.441L...6S,2014PhRvD..89b5017H,2015MNRAS.451.4413R}, decaying dark matter \citep[e.g.,][]{2014MNRAS.445..614W}, and axion dark matter \citep[e.g.,][]{2015MNRAS.451.2479M}. 

From a particle physics perspective, sterile neutrinos of $\sim$ keV mass provide a compelling warm dark matter (WDM) candidate. Produced in the Early Universe by oscillations with the active neutrino\footnote{Sterile neutrinos produced in the decays of other heavier relics, e.g., \cite{2006PhRvL..97x1301K,2006PhLB..639..414S}, typically generate cold dark matter.}, the sterile neutrino is able to generate dark matter of various warmness (or more precisely, suppression in the primordial matter power spectrum) depending on its mass and production mechanism \citep[e.g.,][]{2001PhRvD..64b3501A}. The same mixing responsible for production leads to sterile neutrino decay, which opens a search strategy using X-ray observations from nearby massive dark matter objects \citep{2001ApJ...562..593A}. 

%-----------------------------------------------------------------  
\begin{figure}
\includegraphics[width=80mm,bb = 0 40 520 580]{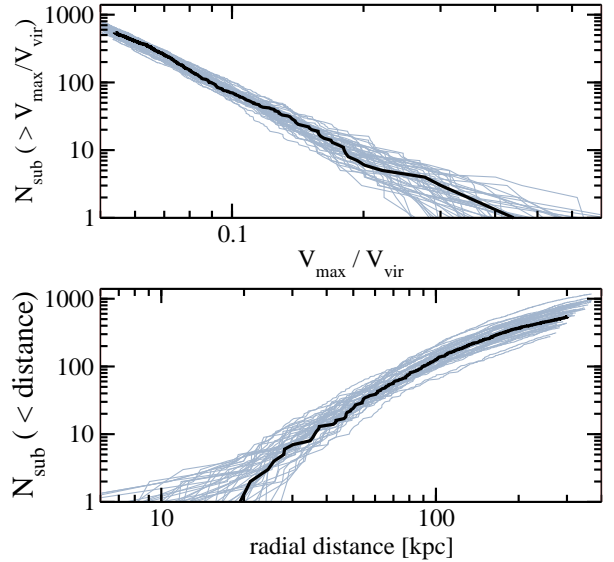}
\caption{Distributions of subhalo $\vmax/\vvir$ (top panel) and radial distance (bottom panel) for the halos simulated in the ELVIS suite (gray). The particular halo used in this study for WDM simulations is highlighted by the thick solid line, indicating its central behavior among the CDM realizations of the ELVIS suite.}
\label{fig:IC}
\end{figure}
%----------------------------------------------------------------

Recently, excess X-ray flux have been observed at ${\sim 3.55}$ keV from observations of stacked galaxy clusters \citep{2014ApJ...789...13B}, Perseus \citep{2014ApJ...789...13B,2014PhRvL.113y1301B}, M31 \citep{2014PhRvL.113y1301B}, the MW \citep{2014arXiv1408.2503B}, and eight additional galaxy clusters \citep{2015arXiv150805186I}. There is ongoing active debate regarding the significance and interpretation of this signal \citep{2015MNRAS.452.3905A,2014arXiv1405.7943R,2015MNRAS.450.2143J,2014arXiv1411.1759J,2014arXiv1409.4143B,2014PhRvD..90j3506M,2015MNRAS.451.2447U,2015PASJ...67...23T,2015arXiv150805186I}. Interpreted as a sterile neutrino decay signal \citep{2014PhRvL.112p1303A}, the simplest scenario is a $\sim 7.1 $ keV sterile neutrino produced by the Shi-Fuller resonantly-enhanced mixing mechanism \citep{1999PhRvL..82.2832S}. 

The focus of this paper is to quantitatively explore the subhalo properties in a $7$ keV sterile neutrino cosmology using dissipationless simulations. We adopt the specific sterile neutrino parameters suggested by the anomalous 3.55 keV line measurement of \cite{2014ApJ...789...13B,2014PhRvL.113y1301B}. More generally, however, our attempt is to simulate the accurate formation of structure in the Shi-Fuller resonant sterile neutrino scenario. In particular, the resonant production of sterile neutrinos has recently been re-investigated by \cite{2015arXiv150706655V}. By including previously neglected effects of the redistribution of lepton asymmetry and the neutrino opacity, as well as a more accurate treatment of the scattering rates through the quark-hadron transition, the authors provide accurate sterile neutrino phase-space densities. Using these updated distributions as inputs, we run dark matter only collisionless $N$-body simulations, which allows us to explore the implications for Local Group satellite counts and their internal kinematics, specifically addressing the missing satellites and TBTF problems. 

The paper is organized as follows. In Section \ref{sec:simulations}, we summarize our simulation and analysis methods. In Section \ref{sec:analysis}, we present our results, including satellite counts in Section \ref{sec:analysis:counts}, satellite radial distribution in Section \ref{sec:analysis:radial}, and internal structure (implications for TBTF) in Section \ref{sec:analysis:tbtf}. Finally, we present discussions and conclusions in Section \ref{sec:discussion}.

%%%%%%%%%%%%%%%%%%%%%%%%%%%%%%%%%%%%%%%%%%%%%%%%
\section{Simulations} \label{sec:simulations}
%%%%%%%%%%%%%%%%%%%%%%%%%%%%%%%%%%%%%%%%%%%%%%%%

%-----------------------------------------------------------------  
\begin{figure}
\includegraphics[width=80mm,bb = 0 40 520 580]{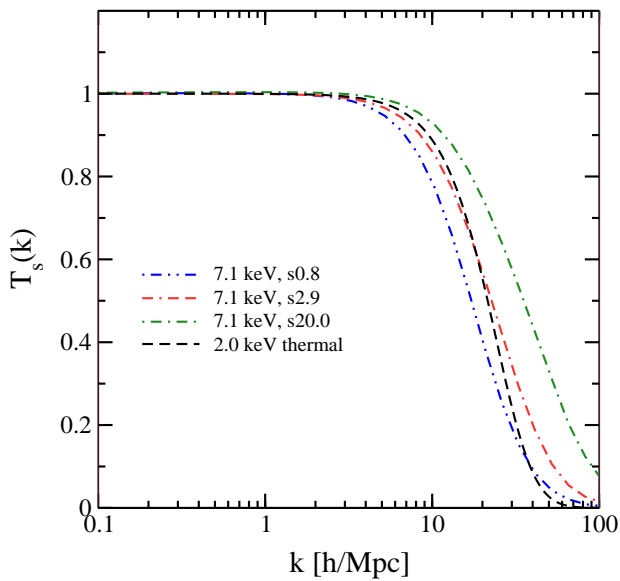}
\caption{Square root of the sterile neutrino power spectrum relative to CDM, $T_s(k) = \sqrt{P_s(k)/P_{\rm CDM}(k)}$. Shown are three resonantly produced sterile neutrino cases, s0.8 (blue dot-dot-dashed), s2.9 (red dot-dashed-dashed), and s20.0 (green dot-dashed). For comparison, a 2.0 keV thermal WDM is shown (black dashed). The initial cutoff shape is similar between the 2.0 keV thermal and s2.9 resonant cases, but the tails show quantitative differences.}
\label{fig:transfer}
\end{figure}
%----------------------------------------------------------------

We run collisionless dark matter-only simulations with the publicly available Tree-PM $N$-body simulation code \texttt{GADGET-2}\footnote{http://www.mpa-garching.mpg.de/gadget/} \citep{2005MNRAS.364.1105S}. We present 4 zoom-in simulations in WDM, all with WMAP7 cosmological parameters: $\sigma_8 = 0.801$, $\Omega_m = 0.266$, $\Omega_b = 0.0449$, $\Omega_\Lambda = 0.734$, $n_s = 0.963$, and $h = 0.71$ \citep{2011ApJS..192...16L}. Simulations were initialized at redshift $125$ with initial conditions selected from the Exploring the Local Volume In Simulations project \cite[ELVIS\footnote{http://localgroup.ps.uci.edu/elvis/index.html};][]{2014MNRAS.438.2578G} and created with {\tt MUSIC}\footnote{http://www.phys.ethz.ch/~hahn/MUSIC/} \citep{2011MNRAS.415.2101H}. ELVIS is a suite of $48$ \lcdm\ zoom-in simulations designed to study the Local Group. It consists of $24$ halos in paired systems that are chosen to resemble the MW and M31 in mass, relative kinematics, and environment, as well as an additional $24$ halos that are isolated mass-matched analogues. Lagrange volumes are determined by all particles within $5 \rvir$ of the halo center in the final timestep for isolated analogues, and $4 \rvir$ of either MW or M31 hosts for pairs. We refer the reader to \cite{2014MNRAS.437.1894O} and \cite{2014MNRAS.438.2578G} for the detail methodology of the zoom-in simulations and ELVIS, respectively. 

We select from the ELVIS suite one halo that is closest to the average over the ELVIS suite in two key subhalo properties -- the subhalo $V_{\rm max}$ and radial position distributions. Here, the circular velocity is $\vcirc = \sqrt{GM(<r)/r}$, and its peak circular velocity and the radius at which it peaks are denoted as $\vmax$ and $\rmax$, respectively. Figure \ref{fig:IC} shows our adopted halo in thick solid black, compared to the entire ELVIS suite (grey solid). 

\begin{table*}
\caption{Properties of the main halo: virial mass $\mvir$, virial radius $\rvir$, circular velocity $\vvir$, peak of the circular velocity $\vmax$, and the radius of peak circular velocity $\rmax$. The last three columns show the number of subhalos within the virial radius with $\vmax > 8$ km/s, $\vmax > 15$ km/s, and $\vpeak > 20$ km/s, respectively. }
\label{table:halo}
\begin{tabular}{@{}lcccccccc}
\hline
		 	&  $\mvir$		& $\rvir$	& $\vvir$	& $\vmax$	& $\rmax$		& Number of subhalos	&  Number of subhalos	&  Number of subhalos  \\
			& ($10^{12} M_\odot$)	&  (${\rm kpc}$)	& (km/s)	& (km/s) 	& ($ {\rm kpc}$) 	& ($\vmax > 8$ km/s) 	& ($\vmax > 15$ km/s) 	& ($\vpeak > 20$ km/s)  \\
\hline
CDM			& 1.56	& 302	& 149	& 179		& 47.7		& 540	& 66		& 57	\\
WDM\_s0.8 	& 1.54	& 301	& 148	& 180		& 46.0		& 35		& 12		& 18	\\
WDM\_s2.9 	& 1.55	& 302	& 149	& 180		& 46.4		& 61		& 20		& 23 \\
WDM\_s20.0	& 1.55	& 301	& 149	& 180		& 45.8		& 105	& 31		& 38 \\
WDM\_2.0keV 	& 1.55	& 301	& 149	& 179		& 48.4		& 48		& 18		& 23	\\
\hline
\end{tabular}
\end{table*}

To build the WDM initial power spectra, CDM transfer functions were first generated using the publicly available \texttt{CAMB}\footnote{http://camb.info/} CMB boltzmann code \citep{2000ApJ...538..473L}. These were modified according to the resonant sterile neutrino production calculations of \cite{2015arXiv150706655V}, which include effects of the redistribution of lepton asymmetry and the neutrino opacity, as well as a more detail treatment during the quark-hadron transition. These affect the primordial momentum distributions of the sterile neutrino, which in turn quantitatively change the matter power spectrum. Figure \ref{fig:transfer} shows the square-root of the suppression of the power spectrum, $T_s(k) = \sqrt{P_s(k)/P_{\rm CDM}(k)}$, for the three sterile neutrino parameters studied in this paper: all have masses of $m_s = 7.1$ keV, but different mixing angles ${\rm sin}^2 2 \theta = 0.800 \times 10^{-11}$ (labeled s0.8), $2.899 \times 10^{-11}$ (labeled s2.9), and $20.000 \times 10^{-11}$ (labeled s20.0). The lepton asymmetries required to reproduce the observed dark matter abundance $\Omega_{\rm dm}$ are $\approx (13.0$--$13.1)\times 10^{-5} $, $(8.32$--$8.39) \times 10^{-5} $, and $(6.7$--$6.8)\times 10^{-5}$, respectively\footnote{These are based on Planck cosmology with $\Omega_{dm}h^2 = 0.119$, but the cosmology dependence is weak \citep{2001ApJ...562..593A}.}. Also shown for comparison is the transfer function for a thermal warm dark matter based on \cite{2001ApJ...556...93B}. Specifically, we use the functional forms in their appendix with values adjusted to our cosmology, and fix the mass to a thermal 2.0 keV WDM particle. The 2.0 keV thermal particle was chosen due to its close resemblance to the s2.9 sterile neutrino, thus serving as a good comparison; it was also identified as being near the center of the range of WDM cutoff scales for the 7.1 keV resonant sterile neutrino in \cite{2014PhRvL.112p1303A}. A companion work by Bozek et al.~2015 simulates a MW/M31 pair from the ELVIS suite in a variety of sterile neutrino models (including the s2.9 model) to examine the impact of these models in the Local Group environment. 

All simulations were run with a $z=0$ Plummer equivalent force softening of $141$ pc in the highest resolution zoom-in region, which contains particles of mass $1.9 \times 10^5 M_\odot$. Resolution tests performed in \cite{2014MNRAS.438.2578G} with $2^3$ more particles (particle mass $2.4 \times 10^4 M_\odot$) show that simulations at our resolution converge for subhalos above $\vmax \gtrsim 8 \, {\rm km/s}$ and $\rmax$ resolved for $\vmax \gtrsim 15 \, {\rm km/s}$. We identify halos with {\tt Rockstar}\footnote{https://code.google.com/p/rockstar/} \citep{2013ApJ...762..109B}, a publicly available six-dimensional friends-of-friends halo finder. Table \ref{table:halo} summarizes the physical properties of the main host halo in the zoom simulations. We define the virial mass $\mvir$ as mass contained within a sphere of radius $\rvir$ that corresponds to an over density of 97 relative to the critical density of the Universe \citep{1998ApJ...495...80B}. We define the $\vmax$ at the moment a halo has its maximum mass as $\vpeak$. Extracting $\vpeak$ thus needs the assembly of a merger tree, which we perform with {\tt consistent-trees}\footnote{https://bitbucket.org/pbehroozi/consistent-trees} \citep{2013ApJ...763...18B}.

A particular concern for WDM $N$-body simulations is the artificial fragmentation of filaments that cause the production of artificial clumps. These act as spurious halos and contaminate the subhalo catalogs. \cite{2014MNRAS.439..300L} investigated the contamination based on a series of dissipationless WDM simulations. They adopted thermal WDM masses of 1.5, 1.6, 2.0, and 2.3 keV, i.e., with cutoffs in the range of interest for our study, and find that the spurious subhalos approximately equal the number of genuine subhalos on scales of $\sim 10^7 M_\odot$ (or $\vmax \sim 5$ km/s) for a 2.0 keV WDM and MW-sized halo, and dominate on smaller scales. They test this using high-resolution (particle mass $1.55 \times 10^4 M_\odot$) and low-resolution (particle mass $4.43 \times 10^5 M_\odot$) analogues, i.e., our resolution falls in between their test cases. These suggest our resolved subhalos (i.e., with $\vmax > 8$ km/s) are not severely contaminated. This is also consistent with the scaling presented by \cite{2007MNRAS.380...93W}. However, the precise form of the subhalo mass function on scales smaller than the cutoff scale is uncertain \citep{2013MNRAS.433.1573S,2013MNRAS.434.1171H}. Instead of attempting to remove spurious subhalos, we avoid the issue by only considering objects larger than $\vmax > 8$ km/s.

%%%%%%%%%%%%%%%%%%%%%%%%%%%%%%%%%%%%%%%%%%%%%%%%
\section{Analyses} \label{sec:analysis}
%%%%%%%%%%%%%%%%%%%%%%%%%%%%%%%%%%%%%%%%%%%%%%%%

%%%%%%%%%%%%%%%%%%%%%%%%%%%%%%%%%%%%%%%%%%%%%%%%
\subsection{Subhalo abundance} \label{sec:analysis:counts}
%%%%%%%%%%%%%%%%%%%%%%%%%%%%%%%%%%%%%%%%%%%%%%%%

\begin{figure}
\includegraphics[width=80mm,bb = 0 40 520 580]{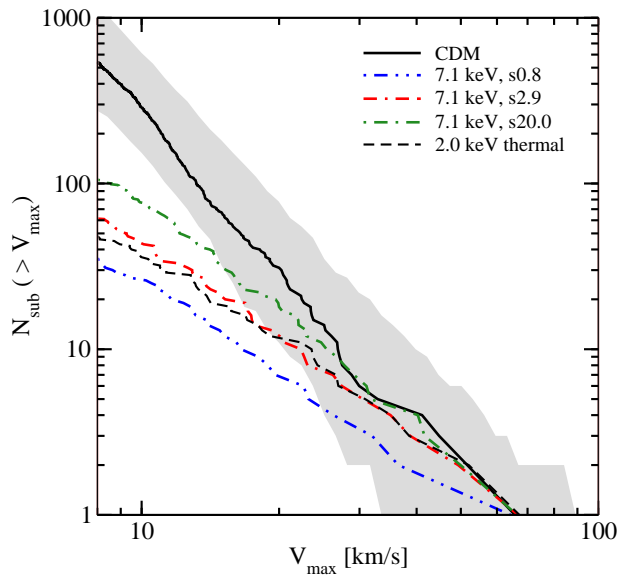}
\caption{Cumulative distributions of subhalos within the virial radius in $\vmax$. Only subhalos with $\vmax > 8$ km/s are shown, due to resolution as well as potential contamination from spurious subhalos at lower $\vmax$. Shown are resonantly generated sterile neutrinos s0.8 (blue dot-dot-dashed), s2.9 (red dot-dashed-dashed), and s20.0 (green dot-dashed), a 2.0 keV thermal WDM (black dashed), and CDM (black solid), as labeled. For the CDM, the gray band indicates the scatter determined from the ELVIS suite of simulations. }
\label{fig:vmax}
\end{figure}

We begin by exploring the abundance of dark matter substructures within the virial radius of the host halo. The three final columns of Table \ref{table:halo} show the total number of subhalos within the virial radius with $\vmax > 8$ km/s, $\vmax > 15$ km/s, and $\vpeak > 20$ km/s cuts, respectively. The cumulative distributions are shown in Figure \ref{fig:vmax}. They reveal a strong suppression in the number of subhalos. The suppression in the number of low-mass subhalos is consistent with the cutoff in the WDM transfer function. For example, s20.0 has the largest cutoff wavenumber (Figure \ref{fig:transfer}) and as a result shows the least suppression. 

According to the standard paradigm, these subhalos must host the satellites orbiting the main galaxy. Observationally, the number of satellites orbiting the MW currently stands at $\sim$28, including 11 so-called classical dwarfs known pre-SDSS, a similar number of ultra-faint dSph galaxies discovered by the SDSS \citep{2010MNRAS.406.1220W,2012AJ....144....4M}, and 8 recently discovered candidates in the DES fields \citep{2015ApJ...807...50B}. We remove from this list Leo T and Eridanus II, which are both clearly further away than the virial radius of the MW. We keep the Large Magellanic Cloud (LMC) and the Small Magellanic Cloud (SMC), since the ELVIS halos were intentionally chosen to contain realistic analogues to the Magellanic clouds \citep{2014MNRAS.438.2578G}. Our chosen halo has one subhalo with $\vmax > 60$ km/s; had halos been chosen at random, it would have been unlikely to find such massive subhalos \citep{2010MNRAS.406..896B}. The satellite list should be taken as a lower limit, since on-going and future more complete and deeper surveys are expected to lead to more discoveries. Indeed, 6 to 8 candidate ultrafaint dSphs were recently reported in the DES field \citep{2015arXiv150803622T}, and the intrinsic satellite count of the Milky Way is estimated to be close to $\sim 100$ \citep{2008ApJ...688..277T,2009AJ....137..450W,2014ApJ...795L..13H,2015arXiv150803622T}. Therefore, the simulations should contain at a minimum 28 subhalos to host the known satellites, and certainly more. 

The equivalent in M31 is aided by the Pan-Andromeda Archeological Survey (PAndAS), which provides complete coverage out to $\sim 150$ kpc from M31 and sensitive to satellites down to luminosities of $\sim 10^5 L_\odot$ \citep{2011ApJ...732...76R}. Larger distances have been subject to various studies, most recently the Panoramic Survey Telescope and Rapid Response System (Pan-STARRS) that provides a survey going deeper than SDSS and wider than PAndAS. The total number of satellites within 150 kpc (300 kpc) is now 18 (35) \citep{2012AJ....144....4M,2012ApJ...758...11C,2013ApJ...772...15M}. While the total M31 satellite population is surprisingly similar to that of the MW, the MW count includes faint satellites ($L < 10^5 L_\odot$) that are currently difficult to detect around M31. In fact, M31 contains significantly more bright satellites than the MW. As in the MW, therefore, current counts should be treated as a lower limit, particularly at the faint end.

To compare, we consider subhalos with $\vpeak> 20$ km/s, because subhalos with lower $\vpeak$ may not have formed stars due to suppressed gas accretion during the reionization epoch \citep{2000ApJ...539..517B,2015MNRAS.448.2941S}. This condition is also consistent with extrapolation of abundance matching to MW satellite scales \citep{2014MNRAS.444..222G}, in particular given the uncertainty in the scaling between stellar and halo masses (Garrison-Kimmel et al., in prep). Comparing the final column in Table \ref{table:halo} to the known satellites, we see that s0.8 and s2.9 are disfavored, while s20.0 satisfies the requirements. However, there is a factor of $\sim 2$ intrinsic spread from main halo selection (grey band in Figure \ref{fig:vmax}). We thus conclude that \emph{all the WDM simulations contain sufficient numbers of subhalos to host the presently observed satellites, but the space for additional satellites is limited.} We caution however that the constraint can be relaxed by considering galaxy formation in subhalos with $\vpeak<20$ km/s. For example, recent simulations suggest galaxy formation can occur in $\vpeak < 20$ km/s subhalos provided the process occurs at high enough redshifts \citep{2015MNRAS.453.1305W}. 

%%%%%%%%%%%%%%%%%%%%%%%%%%%%%%%%%%%%%%%%%%%%%%%%
\subsection{Radial profiles}\label{sec:analysis:radial}
%%%%%%%%%%%%%%%%%%%%%%%%%%%%%%%%%%%%%%%%%%%%%%%%

\begin{figure}
\includegraphics[width=80mm,bb = 0 40 520 580]{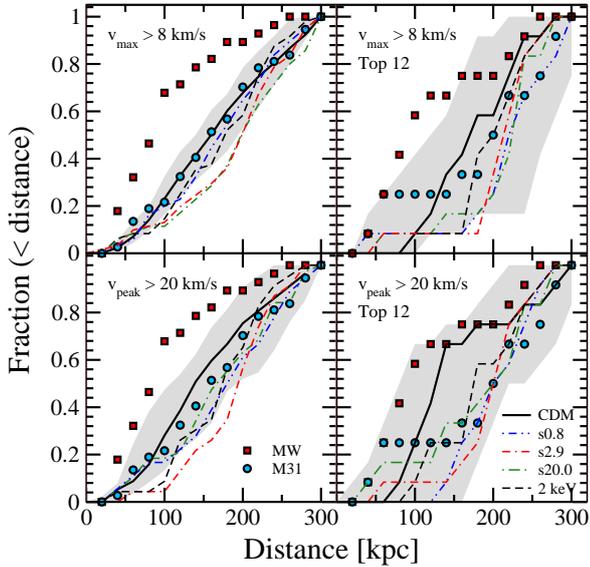}
\caption{Normalized radial distributions of satellites for the MW and M31, and the same for subhalos for five cosmologies: CDM (black solid), thermal 2.0 keV WDM (black dashed), and three resonant sterile neutrino WDM s0.8 (blue dot-dot-dashed), s2.9 (red dot-dashed-dashed), and s20.0 (green dot-dashed), as labeled. The gray band shows the spread in the CDM results based on the ELVIS suite of simulations. The 4 panels show: all satellites and all subhalos with $\vmax > 8$ km/s (top left), all satellites and all subhalos with $\vpeak > 20$ km/s (bottom left), and their equivalents using only the most luminous 12 satellites and the largest 12 subhalos (right panels).}
\label{fig:radial}
\end{figure}

We next explore the radial distributions of the subhalos and compare them to the observed satellites of the MW and M31. To make a comparison of the radial distribution shape, we normalize the distributions by the total number within the virial radius. The left panels of Figure \ref{fig:radial} show the results when all resolved subhalos are compared against the entire known satellites population; the top and bottom panels show results applying a $\vmax > 8$ km/s cut and a $\vpeak > 20$ km/s cut to the subhalo, respectively. In both cases, the simulations are in better agreement with the satellites of M31 than the satellites of the MW, mirroring analyses of the $\Lambda$CDM ELVIS suite \citep{2014MNRAS.439...73Y}. The different cutoffs do not change this result. If additional satellites are discovered at large distances ($> 150$ kpc) around the MW, the radial distribution of MW satellites will more closely match those of simulated subhalos and M31 satellites. On the other hand, there is presently a stronger luminosity bias in the M31 satellites, and future discoveries of dim satellite dwarfs may change the M31 curve. 

%-----------------------------------------------------------------  
%----------------------------------------------------------------

%-----------------------------------------------------------------  
%----------------------------------------------------------------

The right panels show the comparison when only the top 12 most massive subhalos are considered: the top panel uses the 12 largest in $\vmax$, while the bottom panel uses the 12 largest in $\vpeak$. These are compared against the most luminous 12 satellites which are likely to be observationally complete. For the MW, this is the LMC, SMC, Sagittarius, Fornax, Leo I, Sculptor, Carina, Leo II, Sextans, Ursa Minor, Draco, and Canes Venatici I; for M31, Triangulum (M33), And XXXI, NGC 205, M32, IC 1613, IC 10, NGC 185, NGC 147, And VII, And II, And I, and And VI. Due to the smaller numbers, the top 12 subhalo radial distribution shows considerably more scatter than the total. Within the scatter, the radial distributions of the 12 most massive subhalos are marginally consistent with the brightest MW and M31 satellites. 

In all comparisons, WDM realizations tend to yield radially expanded subhalo distributions compared to CDM. This is consistent with the later formation times of subhalos in WDM than in CDM. The later formation times yield not only less concentrated subhalos, but they are also more susceptible to tidal disruption during near passages in the inner halo \citep[e.g.,][]{Maccio':2009dx}. Similar effects are seen in studies by \cite{2012JCAP...10..047A} who explored the radial profiles of subhalos in various WDM and CDM + WDM cosmologies. In general, this makes WDM realizations less compatible with the distribution of MW satellite radial positions, which are more concentrated. Currently, there is only a mild tension when limiting the comparison to the observationally complete luminous satellite sample (right panels of Figure \ref{fig:radial}). However, the radial position distribution will become a powerful diagnostic when a larger sample of complete satellite galaxies becomes available, e.g., from ongoing surveys such as DES, and future missions such as the Large Synoptic Survey Telescope (LSST) and the Wide-Field Infrared Survey Telescope (WFIRST) \citep{2015arXiv150303757S} which will in particular be able to study satellite populations in nearby galaxies out to several Mpc.

We conclude that the radial position distributions of the subhalos in our resonant sterile neutrino realizations are in agreement with the observed satellites of M31. While there remains some tension when compared with the satellites of the MW, this is significantly reduced when only the most luminous classical satellites are considered. Future satellite galaxy samples, both of MW and other galaxies, will help test this further. 

%%%%%%%%%%%%%%%%%%%%%%%%%%%%%%%%%%%%%%%%%%%%%%%%
\subsection{Internal structure of subhalos}\label{sec:analysis:tbtf}
%%%%%%%%%%%%%%%%%%%%%%%%%%%%%%%%%%%%%%%%%%%%%%%%

We finally explore the internal structures of subhalos. The simulations performed in this study do not fully resolve density profiles in the innermost $\sim 500$ pc, but integral values such as $\vmax$ and $\rmax$ are converged for subhalos with $\vmax > 15$ km/s \citep{2014MNRAS.438.2578G}. With these parameters, the two-parameter Navarro-Frenk-White \citep[][NFW]{1997ApJ...490..493N} density profile is fully defined,
\begin{equation}
\rho(r) = \frac{\rho_s}{(r/r_s)(1+r/r_s)^2}, 
\end{equation}
where the characteristic scale radius $r_s = \rmax / 2.1626$ and the scale density $\rho_s = \rho_s(\vmax,\rmax)$. 

%-----------------------------------------------------------------  
\begin{figure}
\includegraphics[width=80mm,bb = 0 40 520 580]{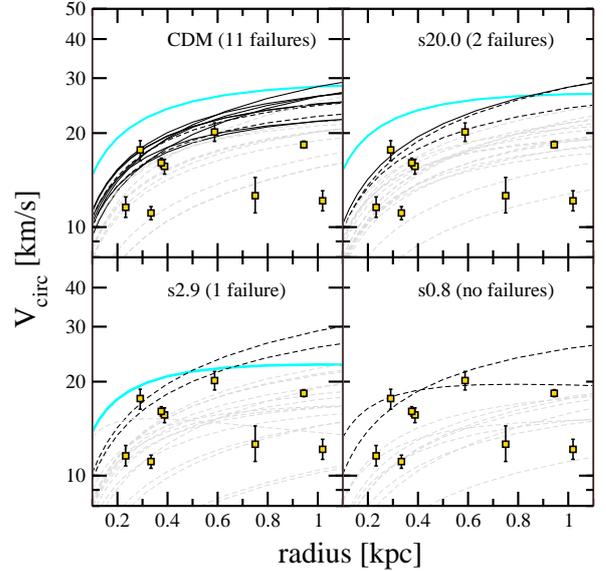}
\caption{Rotation curves of all or 30 of the most massive subhalos with $15<\vmax/({\rm km/s})<60$. Values from measured circular velocities at half-light radii of the nine classical MW dwarfs, excluding LMC, SMC, and Sagittarius, are also plotted (symbols). Massive failures that cannot host any of the MW dSphs in the sample (thick solid cyan), additional massive failures from consideration of Draco and Ursa Minor (thin solid black), subhalos that host either Draco or Ursa Minor (dashed black; only two allowed) and subhalos that are consistent with at least one of the remaining 7 dSphs in the sample (gray dashed), are shown.}
\label{fig:tbtf}
\end{figure}
%----------------------------------------------------------------

By extrapolating the NFW profile, we determine the rotation curves of subhalos, as shown in Figure \ref{fig:tbtf}. Following \cite{2011MNRAS.415L..40B}, we only include subhalos with $15<\vmax/({\rm km/s})<60$; the LMC, SMC, and Sagittarius are excluded for a consistent comparison. For $\Lambda$CDM (top left panel), we show only the 30 most massive subhalos; for the others, we show all the subhalos that satisfy the $\vmax$ cut. These are compared to the circular velocities of dSphs at their de-projected half-light radius ($r=r_{1/2}$). At this radius, accurate dynamical mass estimates are obtained \citep{2010MNRAS.406.1220W}; the vertical error bars are derived from the $1\sigma$ uncertainty in the mass at the half-light radius. 

We use the ``strong massive failure'' definition from \cite{2014MNRAS.444..222G}, which considers all subhalos that cannot be assigned to host a dwarf galaxy as massive failures, with the caveat that the two densest dwarfs (Draco and Ursa Minor) can only be hosted by a single subhalo each. In $\Lambda$CDM, the number of massive failures is typically $\sim 20$, but with a large scatter from 2 to over 40, depending on the halo mass, cosmology, and other variables. Our chosen halo contains 11 massive failures in $\Lambda$CDM. We find that with our specific resonantly produced sterile neutrino dark matter, the number of massive failures is greatly reduced to between none to two. 

An implicit assumption we have made is that the density profiles of subhalos do not deviate from NFW at small radii in a WDM cosmology. This is not valid for radii smaller than some critical value, since WDM predicts the formation of a core beyond a WDM model-dependent density. However, the scales for core formation is less than \emph{O}(10) pc \citep{2011JCAP...03..024V,2012MNRAS.424.1105M,2013MNRAS.428.3715M}, i.e., not an important effect on scales of interest in this study. 

%%%%%%%%%%%%%%%%%%%%%%%%%%%%%%%%%%%%%%%%%%%%%%%%
\section{Conclusions}\label{sec:discussion}
%%%%%%%%%%%%%%%%%%%%%%%%%%%%%%%%%%%%%%%%%%%%%%%%

In this paper, we have tested the near-field cosmological implications of a 7.1 keV sterile neutrino as the origin of the anomalous 3.55 keV lines recently discovered towards a number of massive dark matter objects. Specifically, we have simulated structure formation of a MW-size host from $z=125$ to $z=0$ using the latest primordial matter perturbation power spectrum implied by sterile neutrinos generated via the Shi-Fuller resonance mechanism, and explored the properties of their subhalos: their number, radial distribution, and internal structure. We explored 3 sterile neutrino mixing angles, all consistent with the observed 3.55 keV line and reproducing the cosmological dark matter abundance, but different in their power spectrum cutoff (Figure \ref{fig:transfer}). We also simulate a 2.0 keV thermal WDM candidate for comparison purposes. 

We find that a sterile neutrino responsible for the 3.55 keV line implies non-trivial differences in the subhalo properties of a MW-sized halo compared to CDM and thermal WDM. The number of subhalos available to host satellite galaxies is reduced compared to CDM by factors of $\sim 2$--3 (Table \ref{table:halo}). The suppressions are already significant: in some cases, the predicted number of subhalos is already smaller than the number of observed satellites of the MW and M31. The sterile neutrino models are not ruled out however, once scatter is included, e.g., arising owing to the uncertain mass of the MW halo (Figure \ref{fig:vmax}). The distributions of the subhalo radial positions are systematically less concentrated than the observed satellite positions, although they are consistent when only the observationally-complete luminous satellites and most massive subhalos are considered (Figure \ref{fig:radial} bottom right panel). Finally, the internal properties of the most massive subhalos show a consistent reduction in the number of strong massive failures (Figure \ref{fig:tbtf}). 

%-----------------------------------------------------------------  
\begin{figure}
\includegraphics[width=80mm,bb = 0 40 520 580]{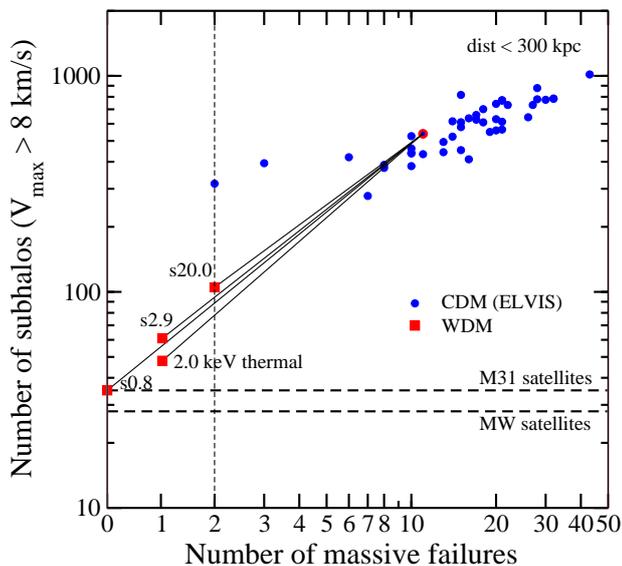}
\caption{Summary of CDM and WDM zoom-in simulation results, showing the number of all subhalos within 300 kpc of the host and with $\vmax > 8$ km/s (y-axis), and the number of massive failures (x-axis; note the axis transitions from linear to logarithmic scale at 2 massive failures indicated by the vertical dashed line). Blue circles indicate CDM ELVIS simulations, and the halo selected to run in WDM highlighted by a red circle. The red squares indicate WDM with differing WDM particles, as labeled. 
}
\label{fig:summary}
\end{figure}
%----------------------------------------------------------------

The fact that the subhalo counts are very close to the number of satellites has two immediate implications. Firstly, future discoveries of satellites will provide strong and direct tests of the resonantly produced sterile neutrino scenario. Indeed, the recent report of 6--8 candidate ultrafaint dwarf satellites \citep{2015arXiv150803622T} suggests a total of $\sim 100$ such objects should be found around the Milky Way, consistent with theoretical predictions \citep{2008ApJ...688..277T,2014ApJ...795L..13H}. This would, e.g., be a factor $\sim 3$ larger than the resolved subhalo counts of the s0.8 scenario and put the s0.8 interpretation of the 3.55 keV in serious tension with observations. 

Secondly, the closeness highlights the future importance of the details of the cutoff shape. In our example, the difference between s2.9 and 2.0 keV thermal models are minimal for the most massive subhalos ($\vmax > 20$ km/s), but the 2.0 keV model gradually shows less subhalos than s2.9 at smaller masses. The difference in the number of resolved ($\vmax > 8$ km/s) subhalos is $13$, which is a 20--30\% effect. The smaller number of subhalos appears consistent with the stronger cutoff in the 2.0 keV model with respect to the s2.9 model (Figure \ref{fig:transfer}). However, the difference is most stark near the resolution limit, and another major concern is the degree of contamination of spurious halos. These make it difficult to provide a robust quantitative statement. Nevertheless, our results indicate that studies of small-scale structures of sterile neutrino WDM is becoming sensitive to the details of the cutoff shape, and thus the exact resonant sterile neutrino cutoff should be used, rather than the approximate thermal equivalent. The impacts of the cutoff will be investigated, in a companion work, in the context of a wider range of transfer function shapes both in the Local Group and in representative volumes of the Universe at high redshift (Bozek et al. 2015).

Figure \ref{fig:summary} shows a summary of our findings, plotting the number of subhalos within 300 kpc and with $ \vmax > 8$ km/s on the $y$-axis, and the number of strong massive failures (based on  $15 < \vmax / ({\rm km/s}) < 60$ subhalos) on the $x$-axis. The blue circles denote the results based on the $\Lambda$CDM simulations of the ELVIS suite, with the specific halo simulated in WDM shown with a red perimeter. The red squares are our WDM simulations, both the resonantly produced sterile neutrinos and thermal WDM, as labeled. The observed satellite counts of the MW and M31 are shown by dashed lines and labeled. The missing satellites problem is illustrated by the $y$-axis: the CDM simulations (blue circles) all produce far too many subhalos relative to observed satellite galaxy counts (dashed horizontal lines). The $x$-axis represents TBTF: the CDM simulations predict a large population of dense subhalos that cannot host the known MW dSphs. Sterile neutrino dark matter is effective at reducing the severity of the TBTF problem while matching the subhalo abundance. Overall, we conclude that a 7.1 keV sterile neutrino provides a good description of the Local Group, which is often better than CDM in dissipationless simulations, but will be tested by future searches for MW and M31 satellites.

%%%%%%%%%%%%%%%%%%%%%%%%%%%%%%%%%%%%%%%%%%%%%%%%
\section*{Acknowledgments}
%%%%%%%%%%%%%%%%%%%%%%%%%%%%%%%%%%%%%%%%%%%%%%%%

We thank Tejaswi Venumadhav and Francis-Yan Cyr-Racine for discussions and for providing digitized data of sterile neutrino transfer functions. S.~H.~and K.~N.~A.~acknowledge support from the Institute for Nuclear Theory program ``Neutrino Astrophysics and Fundamental Properties'' 15-2a where part of this work was done. K.~N.~A.~is partially supported by NSF CAREER Grant No.~PHY-1159224 and NSF Grant No.~PHY-1316792. MBK acknowledges support provided by NASA through HST theory grants (programs AR-12836 and AR-13888) from the Space Telescope Science Institute (STScI), which is operated by the Association of Universities for Research in Astronomy (AURA), Inc., under NASA contract NAS5-26555. Support for S.G.K.~was provided by NASA through Einstein Postdoctoral Fellowship grant number PF5-160136 awarded by the Chandra X-ray Center, which is operated by the Smithsonian Astrophysical Observatory for NASA under contract NAS8-03060.

%%%%%%%%%%%%%%%%%%%%%%%%%%%%%%%%%%%%%%%%%%%%%%%%%%

%%%%%%%%%%%%%%%%%%%% REFERENCES %%%%%%%%%%%%%%%%%%

% The best way to enter references is to use BibTeX:

\bibliographystyle{mnras}
\bibliography{wdm} % if your bibtex file is called example.bib

% Alternatively you could enter them by hand, like this:
% This method is tedious and prone to error if you have lots of references
%\begin{thebibliography}{99}
%\bibitem[\protect\citeauthoryear{Author}{2012}]{Author2012}
%Author A.~N., 2013, Journal of Improbable Astronomy, 1, 1
%\bibitem[\protect\citeauthoryear{Others}{2013}]{Others2013}
%Others S., 2012, Journal of Interesting Stuff, 17, 198
%\end{thebibliography}

%%%%%%%%%%%%%%%%%%%%%%%%%%%%%%%%%%%%%%%%%%%%%%%%%%

%%%%%%%%%%%%%%%%% APPENDICES %%%%%%%%%%%%%%%%%%%%%

%\appendix

%\section{Some extra material}

%If you want to present additional material which would interrupt the flow of the main paper,
%it can be placed in an Appendix which appears after the list of references.

%%%%%%%%%%%%%%%%%%%%%%%%%%%%%%%%%%%%%%%%%%%%%%%%%%

% Don't change these lines
\bsp	% typesetting comment
\label{lastpage}
\end{document}